\documentclass[3p,times,twocolumn]{elsarticle}

\usepackage{ecrc}

\usepackage{graphicx,subfigure}
\usepackage{amsmath,amssymb,amsfonts}
\usepackage{bm}
\usepackage{color}
\usepackage{comment}
\usepackage{slashed} 
\usepackage{lineno}

\def \be {\begin{equation} }
\def \ee {\end{equation}}
\def \bea {\begin{eqnarray}}
\def \eea {\end{eqnarray}}
\def \bem {\begin{multline}}
\def \eem {\end{multline}}

\def \bes {\begin{subequations} }
\def \ees {\end{subequations}}

\def \x {\xi}

\def \<{\langle}
\def \>{\rangle}
\def \+{\dagger}
\def \({\left(}
\def \){\right)}

\def \[{\left[}
\def \]{\right]}

\newcommand{\bq}{{\bm{q}}}

\usepackage[normalem]{ulem}  

\renewcommand\sout{\bgroup \color{red} \ULdepth=-.5ex \ULset}

\graphicspath{{./figs/}}

\volume{00}

\firstpage{1}

\journalname{Nuclear and Particle Physics Proceedings}

\runauth{}

\jid{nppp}

\jnltitlelogo{Nuclear and Particle Physics Proceedings}

\usepackage{amssymb}

\usepackage[figuresright]{rotating}

\begin{document}

\begin{frontmatter}



\dochead{}

\title{Heavy-Quark Diffusion Dynamics in Quark-Gluon Plasma 
\\
under Strong Magnetic Fields}


\author{Koichi Hattori\fnref{KH}}
\address{Physics Department and Center for Particle Physics and Field Theory, Fudan University, Shanghai 200433, China}

\fntext[KH]{Speaker in the conference.}

\author{Kenji Fukushima}
\address{ Department of Physics, The University of Tokyo, 7-3-1 Bunkyo-ku,
 Hongo, Tokyo 113-0033, Japan}

\author{Ho-Ung Yee}
\address{Department of Physics, University of Illinois, Chicago, Illinois
 60607, U.S.A.}
\address{ RIKEN-BNL Research Center, Brookhaven National Laboratory, Upton,
 New York 11973-5000, U.S.A.}
 
\author{Yi Yin}
\address{Center for Theoretical Physics, Massachusetts Institute of Technology, Cambridge, MA 02139 USA}

\begin{abstract}
We discuss heavy-quark dynamics in the quark-gluon plasma under a strong magnetic field induced by colliding nuclei. By the use of the diagrammatic resummation techniques for Hard Thermal Loop and the external magnetic field, we show analytic results of heavy-quark diffusion coefficient and drag force which become anisotropic due to the preferred spatial orientation in the magnetic field. We argue that the anisotropic diffusion coefficient gives rise to an enhancement/suppression of the heavy-quark elliptic flow depending on the transverse momentum. 
\end{abstract}


\end{frontmatter}

\begin{figure*}[!t]
 \begin{center}
   \includegraphics[clip,width=1.8\columnwidth]{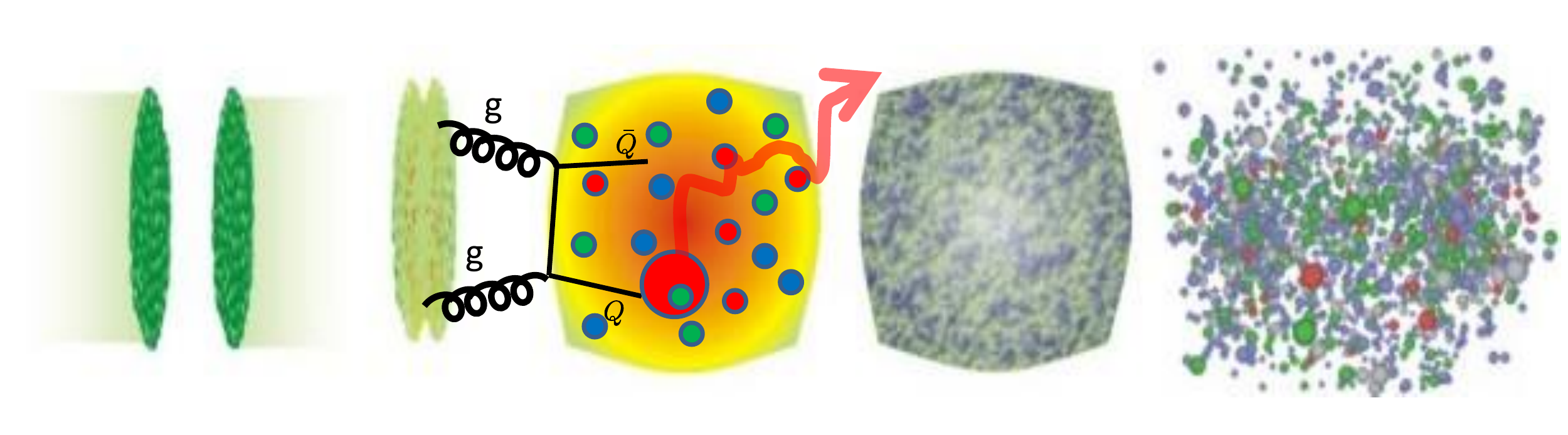}
    \end{center}
    \vspace{-.5cm}
     \caption{Brownian motion of heavy quarks created by the initial hard processes.}
 \label{fig:Brownian}
\end{figure*}

\section{Introduction}
 
In the relativistic heavy-ion collisions, the heavy quarks are dominantly produced 
in the initial hard scatterings among the partons from the colliding nuclei, 
and thus will serve as a probe of the dynamics in the QGP phase and the initial stage. 
This is because the thermal excitation is suppressed due to the large value of
the heavy-quark mass compared to the temperature of the quark-gluon plasma (QGP). 

In the RHIC and LHC experiments, the nuclear modification factor $ R_{AA}$
and the anisotropic spectrum $ v_2$ of the open heavy flavors have been measured. 
Both of them are thought to be sensitive to the heavy-quark thermalization,
and are closely related to each other in the diffusion dynamics \cite{Moore:2004tg}.
However, a consistent theoretical modelling, which simultaneously
reproduces $ R_{AA}$ and $v_2 $, is sill under investigation. 

In this contribution, we discuss effects of an extremely strong magnetic field 
on the heavy-quark diffusion dynamics in QGP. The strong magnetic field is induced 
by the highly accelerated incident nuclei, and its magnitude of the magnetic field 
is estimated to be of the order of pion mass square or even larger (see Ref.~\cite{Hattori:2016emy} and references therein). 
This magnetic field is so strong that the motion of thermal light quarks in QGP 
are strongly modified. Based on the Langevin equation for the Brownian motion of the heavy quarks (see Fig.~\ref{fig:Brownian}), 
we investigate how this modification of the thermal excitation in QGP is reflected 
in the spectra of open heavy flavors \cite{Fukushima:2015wck}. 
The heavy-quark diffusion coefficient is computed by mean of the 
resummation techniques for Hard Thermal Loop and the strong external magnetic field. 
We will find that the heavy-quark diffusion coefficient becomes anisotropic, 
and argue that the anisotropic heavy-quark flow can be generated even before 
the full development of the background fluid flow in the initial stage of the heavy-ion collisions.

\section{Modelling by Langevin equations}

When the heavy quark is subject to the random kicks by the thermal particles, 
the heavy quark dynamics is modelled as a Brownian motion \cite{Moore:2004tg}, 
which is described by the Langevin equations\footnote{
In another contribution talk in the conference \cite{greco, Das:2016cwd}, 
the authors discussed the Lorentz force exerting on the heavy quark 
with the inclusion of the Lorentz force in the Langevin equation (\ref{eq:Langevin}).}: 
\begin{eqnarray}
\label{eq:Langevin}
\frac {d p_z}{ dt}
 = -\eta_\parallel p_z + \xi_z \,,
 \qquad
 {d\bm p_\perp\over dt} =
   -\eta_\perp \, \bm p_\perp + \bm\xi_\perp \,.
\end{eqnarray}
Since the external magnetic field provides a preferred spatial direction,
we have a set of two equations for the heavy-quark motion,
parallel and perpendicular to the magnetic field that is oriented in the $z$-direction. 

The random forces are assumed to be white noises, 
\begin{subequations}
\begin{eqnarray}
&& \hspace{-0.5cm}
 \langle \xi_{z}(t) \xi_z(t')\rangle=\kappa_\parallel \delta(t-t')
 \,,
\\
&& \hspace{-0.5cm}
 \langle \xi_{\perp}^i(t)\xi_\perp^j(t')\rangle=\kappa_\perp
  \delta^{ij}\delta(t-t') \quad (i,j=x,y)
  \, ,
\end{eqnarray}
\end{subequations}
and these coefficients, $\kappa_\parallel$ and $\kappa_\perp$, are
related to the drag coefficients, $\eta_\parallel$ and $\eta_\perp$,
through the fluctuation-dissipation theorem as
\begin{eqnarray}
\label{eq:fdtheorem}
&&
  \eta_{\parallel} = 2 M_{Q}T \kappa_{\parallel} 
\, ,
\quad 
  \eta_{\perp}= 2 M_{Q}T \kappa_{\perp}
\, .
\end{eqnarray}
At the leading order in $ g_s$, the anisotropic momentum diffusion coefficients, $\kappa_\parallel$ and
$\kappa_\perp$, are defined by
\begin{subequations}
\begin{eqnarray}
&&
\kappa_{\parallel} = \int \!\! {d^3\bm q}\,
  {d\Gamma(\bm q)\over d^3\bm q}\, q_{z}^2 \,,
\label{kappa1}
\\
&&
 \kappa_{\perp} = {1\over 2}\int \!\! {d^3\bm q}\,
  {d\Gamma(\bm q)\over d^3\bm q}\, \bm q_{\perp}^2
  \, .
\label{kappa2}
\end{eqnarray}
\end{subequations}
where $ q$ is the amount of the momentum transfer from the thermal particles to the heavy quark,
and the static limit ($ q^0 \to 0$) is assumed in the above definitions. 
In the next section, we show perturbative computation of these transport coefficients 
in the hot medium and the strong magnetic field.

\section{Diffusion coefficients in strong magnetic fields}

\begin{figure}[t!]
		\begin{center}
			\includegraphics[width=0.95\hsize]{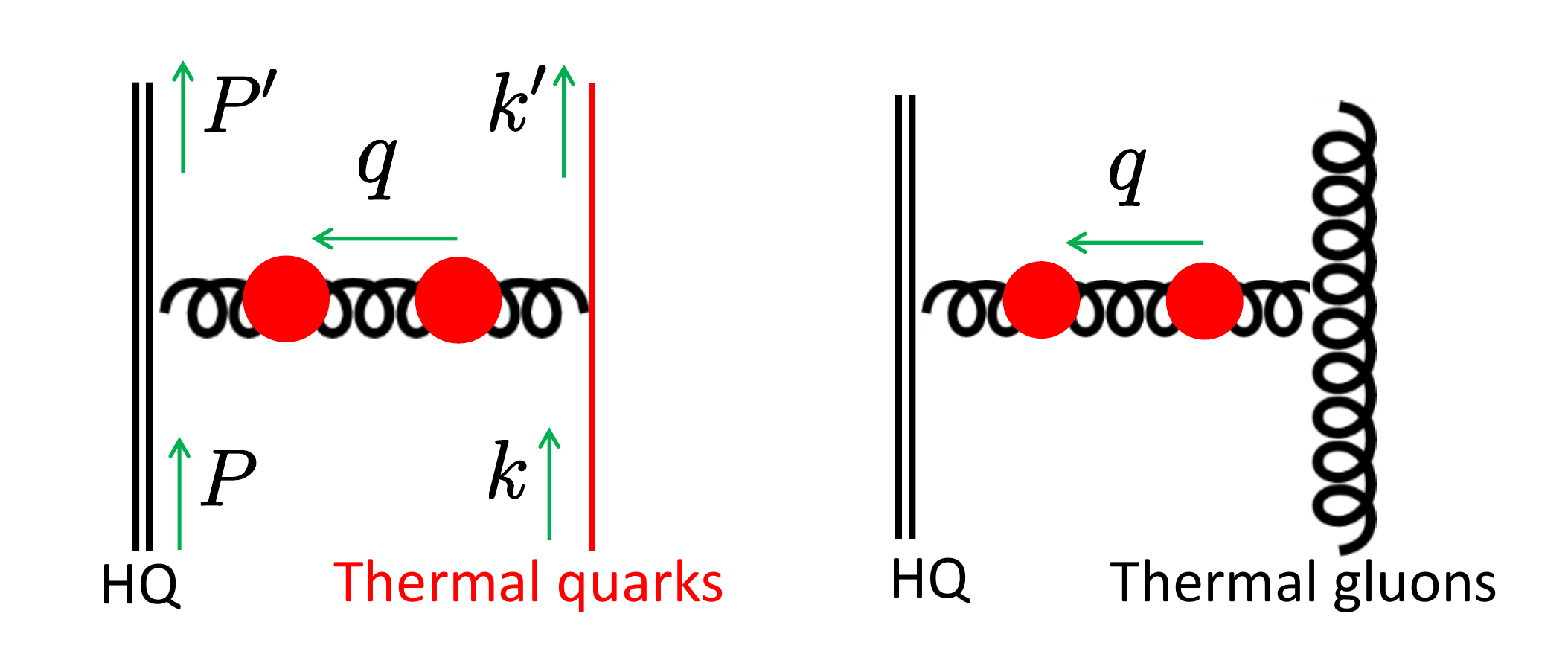}
		\end{center}
        \vspace{-0.5cm}
\caption{Coulomb scattering amplitudes contributing to the heavy-quark momentum diffusion.
A magnetic field acts on the quark loop in the polarization and the thermal-quark scatterers.}
\label{fig:scatterings}
\end{figure}

We compute the diffusion coefficients defined in Eqs.~(\ref{kappa1}) and (\ref{kappa2}). 
At the leading order, contributions to the momentum transfer rate $   {d\Gamma(\bm q)\over d^3\bm q}$ 
come from Coulomb scatterings. 
In the diagrams shown in Fig.~\ref{fig:scatterings}, 
effects of the magnetic field appear in two places (highlighted by red): 
(i) the Debye screening mass and (ii) the dispersion relation of the thermal-quark scatterers. 
On the other hand, the gluons are not directly coupled to the magnetic field, 
so that the modification of their dispersion relation is negligible at the leading order. 

In a strong magnetic field, the fermion wave function is strongly squeezed along the magnetic field,
corresponding to the small cyclotron radius.
Indeed, from the Landau level quantization and the Zeeman effect,
the fermions in the lowest Landau level (LLL) have the (1+1) dimensional dispersion relation, i.e., 
$ \epsilon^2 = m^2 + p_z^2$ for massive fermions and
$ \epsilon = \pm p_z$ for massless fermions with right and left handed chiralities.
To be specific, we focus on the strong-field regime such that
the transition from the LLL to the hLL states are suppressed according to a hierarchy $ T^2 \ll eB $. 

(i) First, we compute the Debye screening mass. In the LLL, 
the gluon self-energy is completely factorized into the transverse and longitudinal parts as \cite{Fukushima:2015wck}
\begin{eqnarray}
\Pi^{\mu\nu} (q)  = q_f \frac{ eB}{2\pi} e^{ - \frac{\vert\bq_\perp\vert^2}{2\vert q_f eB \vert} }
\Pi_{1+1}^{\mu\nu} (q_\parallel)
\, ,
\end{eqnarray}
where the factor of $ q_f \frac{ eB}{2\pi}$ comes from the degeneracy factor
in the transverse phase space, and the Gaussian is the wave function of the LLL state. 
We will find that this quark-loop contribution is much larger than 
the thermal gluon-loop contribution. 
Therefore, the Debye screening mass is obtained form the the (1+1) dimensional quark loop 
which is well-known in the Schwinger model as 
\begin{eqnarray}
\Pi_{1+1}^{\mu\nu} (q_\parallel) = {\rm tr}[t^at^a]
\frac{g_s^2}{\pi}  f(q^0,q_z)
(q_\parallel^2 g_\parallel^{\mu\nu} - q^\mu_\parallel q^\nu_\parallel)
\, .
\end{eqnarray}
Here, the longitudinal momentum is defined by $ q_\parallel^\mu = (q^0,0,0,q_z)$.
Note that there is only one gauge-invariant tensor structure in the (1+1) dimension,
so that the temperature correction is, if any, contained in a scalar function $f(q^0,q_z) $.
However, an important observation is that
there is no temperature or density correction in the massless case 
(see Ref.~\cite{
Fukushima:2015wck} and references therein),
and hence $ f(q^0,q_z) = 1 $.
In massive case, a temperature correction is proportional
to the quark mass, and is suppressed by $m_q/T $ 
(see Ref.~\cite{Fukushima:2015wck} for explicit expressions in the massive case).
Therefore, in both cases, the Debye mass in strong magnetic fields ($eB \gg T^2 $)
is given by the vacuum part of the gluon self-energy \cite{Fukushima:2011nu,Hattori:2012je}.
Namely, we find $ m_D^2 \sim q_f \frac{eB}{2\pi} \cdot \frac{g_s^2}{2\pi} $ 
which is much larger than the usual thermal mass squared $ \sim (g_s T)^2$.

(ii) The change in the dispersion relation of fermion scatterers
is important for the kinematics of the Coulomb scattering process (see Fig.~\ref{fig:scatterings}).
We shall consider the massless fermions in the lowest Landau level \cite{Fukushima:2015wck, Sadofyev:2015tmb}.
For the massless quarks to be on-shell both in the initial and final states,
the energy transfer has to be $q^0 = \pm (k_z^\prime - k_z) $, and thus $ q^0 = \pm q_z$.
Note that the chirality does not change at any perturbative vertex,
so that the signs appear only as the overall ones.
When taking the static limit $q^0 \to 0 $ in the definition of the diffusion coefficients, 
one finds $ q_z \to 0$, and can immediately conclude that the longitudinal-momentum transfer
is kinematically prohibited in the massless case.
On the other hand, this constraint does not apply to the transverse momentum transfer,
so that the transverse momentum transfer is allowed.

The finite contribution to the transverse diffusion coefficient can be obtained either by the direct computation
of the matrix elements in Fig.~\ref{fig:scatterings} or by using the cutting rule.
At the leading order in $ \alpha_s$, the thermal-quark contribution to the momentum diffusion coefficients 
in the massless limit is obtained as \cite{Fukushima:2015wck}
\begin{eqnarray}
 \kappa_\parallel^{\rm LO} = 0 
 \, ,
 \, \quad 
 \kappa_\perp^{\rm LO} \sim \alpha_s^2 T
  \biggl({eB\over 2\pi}\biggr) \ln \alpha_s^{-1}
\label{eq:LO}
\, .
\end{eqnarray}
As discussed above, the quark contribution to the longitudinal diffusion coefficient is vanishing. 
Clearly, the strong magnetic field gives rise to an anisotropy of the momentum diffusion coefficient.

\begin{figure*}
\begin{minipage}[t]{0.49\hsize}
		\begin{center}
			\includegraphics[width=0.8\hsize]{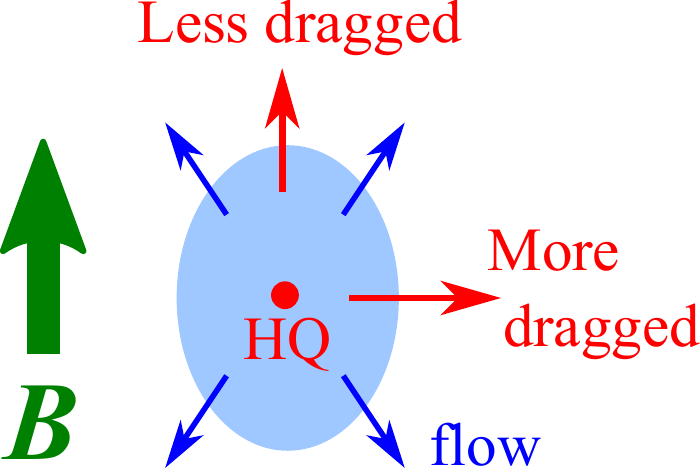}
		\end{center}
		\caption{Schematic pictures of the anisotropic drag force in $ B$.}
		\label{fig:drag}
 \end{minipage}
\begin{minipage}[t]{0.49\hsize}
		\begin{center}
			\includegraphics[width=\hsize]{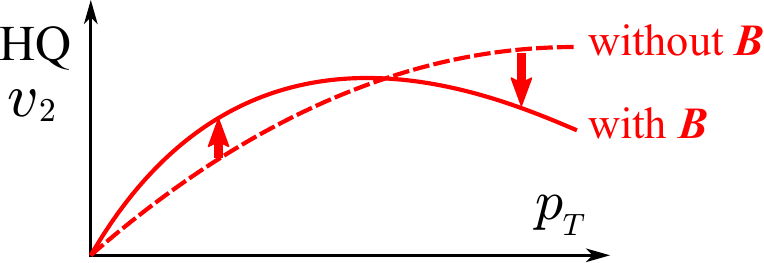}
		\end{center}
\caption{Anisotropic heavy-quark spectrum in the final state.}
\label{fig:v2}
\end{minipage}
\end{figure*}

Nonvanishing contributions to the longitudinal momentum transfer come from
the finite quark-mass correction or the contribution of the gluon scatterers.
The mass correction was obtained as
\begin{eqnarray}
\kappa_\parallel^{\rm LO,\,massive} \sim
\alpha_s    m_q^2 \sqrt{\alpha_s eB}
    \label{eq:LO_mass}
    \, ,
\end{eqnarray}
while the gluon contribution is obtained by substituting the Debye mass $ m_D^2 \sim \alpha_s eB$
for the conventional one $ \sim (g_sT)^2$ in Refs.~\cite{Moore:2004tg} as
\begin{eqnarray}
 \kappa^{\rm LO,\,gluon}_\parallel
\sim \alpha_s^2 T^3 \ln \alpha_s^{-1}
\label{eq:LO_gluon}
  \, .
\end{eqnarray}
We have assumed a hierarchy $ \alpha_s eB \ll T^2 \ll eB$.
Complete expressions including the prefactors are found in Ref.~\cite{Fukushima:2015wck}.
It is instructive to compare the mass correction \eqref{eq:LO_mass} with the gluon contribution \eqref{eq:LO_gluon}.
The ratio is written as
\begin{eqnarray}
\frac{\kappa_\parallel^{\rm LO,\,massive} }{\kappa_\parallel^{\rm LO,\,gluon}} 
&\sim&
\frac{\alpha_s (\alpha_s eB)^{1/2} m_q^2 }{ \alpha_s^2 T^3}
\nonumber
\\
&=& 
\biggl( \frac{m_q^2 }{ \alpha_s eB}\biggr)
 \biggl( \frac{\alpha_s eB }{ T^2}\biggr)^{1/2}
 \biggl( \frac{eB}{T^2}\biggr)
 \,.
\end{eqnarray}
While the first two factors are small in our working regime,
the last factor can be large.  Therefore, the massive-quark contribution
$\kappa_\parallel^{\rm LO,\, massive}$ could be in principle as
comparably large as $\kappa_\parallel^{\rm LO,\, gluon}$, and this
happens when $eB\sim \alpha_s(T^6/m_q^4)$.  However, to be consistent
with our assumed regime, $\alpha_s\,eB\ll T^2$, we have a constraint
of $\alpha_s\ll m_q^2/T^2$, which is not quite likely true in the heavy ion collisions.
Hence,  the longitudinal momentum diffusion coefficient 
is dominated by the gluon contribution $\kappa_\parallel^{\rm LO,\,gluon}$.

Now that we obtained the leading contributions in Eqs.~(\ref{eq:LO}) and (\ref{eq:LO_gluon}),
the anisotropy in the momentum diffusion coefficient and the drag force is estimated to be
\begin{eqnarray}
\frac{\kappa_\parallel^{\rm LO,\,gluon}}{\kappa_\perp^{\rm LO}}
=
\frac{\eta_\parallel^{\rm LO,\,gluon}}{\eta_\perp^{\rm LO}}
\sim \frac{T^2}{eB} \ll 1
\, .
\end{eqnarray}
The quark contribution $ \kappa_\perp^{\rm LO}$ is enhanced by the density of states by the factor of $ eB$. 
On the other hand, a magnetic field does not change the phase space volume of the thermal gluons
in the gluon contribution $ \kappa_\parallel^{\rm LO,\,gluon}$, but changes only the Debye screening mass. 
Summarizing, we have found that the large anisotropy of the diffusion coefficient is induced by 
the prohibition of the longitudinal momentum transfer and 
the enhancement of the phase space both arising in the thermal-quark contribution. 

\section{Summary and discussion}

In summary, we found a large anisotropy of the diffusion coefficients 
induced by a strong magnetic field $ eB \gg T^2$. 
The key observation was the kinematical constraint 
for the momentum transfer in the LLL, which prohibits 
the longitudinal momentum transfer in the massless limit \cite{Fukushima:2015wck, Sadofyev:2015tmb}. 

Based on these findings, 
a toy model for the time evolution of the heavy quark in QGP was discussed in Ref.~\cite{Fukushima:2015wck}
by using the Fokker-Planck equation \cite{Moore:2004tg}.
Since the friction exerting on the heavy quarks is stronger
in the direction transverse to the magnetic field (see Fig.~\ref{fig:drag}),
the heavy-quark momentum in the final state will acquire an anisotropy 
as illustrated in Fig.~\ref{fig:v2}.
In the small momentum region, 
the heavy quarks will be dragged by the flow more strongly in the transverse direction,
while they will feel a larger resistance in the transverse direction
as the momentum increases.
Therefore, there is a turnover in the anisotropy
of the heavy quark spectrum (see the right panel in Fig.~\ref{fig:drag}). 
We leave implementation of the anisotropic diffusion and drag coefficients 
in numerical simulations as a future work. 

Lastly, the techniques and ingredients developed in this study 
can be applied to the computation of other transport coefficients. 
After Ref.~\cite{Fukushima:2015wck}, a few works have been done 
for the jet quenching parameter \cite{Li:2016bbh} 
and the electrical conductivity \cite{Hattori:2016lqx,Hattori:2016cnt} 
in the strong magnetic fields.

\vspace{0.5cm}

Acknowledgement.---This work is supported by 
the 1000 Young Talents Program of China and 
China Postdoctoral Science Foundation under Grant No. 2016M590312 (KH). 

\bibliographystyle{elsarticle-num}



\end{document}